**CARS polarized microscopy of three-dimensional director structures in liquid crystals**

A.V. Kachynski[1], A.N. Kuzmin[1], P.N. Prasad[1,a], and I.I. Smalyukh[2,b]

[1]*The Institute for Lasers, Photonics, and Biophotonics, University at Buffalo, The State University of New York, Buffalo, NY 14260, USA*
[2]*Department of Physics and Liquid Crystal Materials Research Center, University of Colorado at Boulder, Boulder, CO 80309, USA*

We demonstrate three-dimensional vibrational imaging of director structures in liquid crystals using coherent anti-Stokes Raman scattering (CARS) polarized microscopy. Spatial mapping of the structures is based on sensitivity of a polarized CARS signal to orientation of anisotropic molecules in liquid crystals. As an example, we study structures in a smectic material and demonstrate that single-scan CARS and two-photon fluorescence images of molecular orientation patterns are consistent with each other and with the structure model.

Long-range orientational order is an important property of liquid crystals (LCs) [1]. Local average molecular orientations are described by the director $\hat{n} \equiv -\hat{n}$, which is an optical axis in the uniaxial LC materials. Non-invasive imaging of the three-dimensional (3-D) spatial patterns of $\hat{n}(\vec{r})$ is important for fundamental LC research and for a broad range of technological applications. Fluorescence Confocal Polarizing Microscopy (FCPM) [2] visualizes 3-D director fields by taking advantage of (a) doping anisometric dyes that homogeneously distribute and align in the LC and (b) polarized excitation and fluorescence detection. In this approach, the absorption/emission transition dipoles of elongated dye molecules follow $\hat{n}(\vec{r})$ and the polarized confocal imaging visualizes the 3-D director patterns [2]. However, this approach requires doping a specially selected dye that at small concentrations would provide strong contrast without affecting the LC properties. The labeling-free technique of interest is confocal Raman microscopy [3,4], which utilizes a Raman vibration and the signal dependence on orientations of molecular bonds. To study 3-D molecular orientation patterns, several techniques employ nonlinear processes, such as second harmonic generation (SHG) [5], third harmonic generation (THG) [6], and two-photon fluorescence (TPF) [7]. Coherent anti-Stokes Raman scattering (CARS) microscopy has been used for visualizing chemical composition in biological and lyotropic systems [8,9]. We report 3-D imaging of LC director structures using CARS polarized microscopy (CARSPM), which does not require doping the LC with dyes and offers chemical selectivity and bond-orientation specificity superior to that of other techniques [2-7]; it also provides $\sim 10^5$ times faster imaging (stronger signal) than confocal Raman microscopy [3,4]. We demonstrate that CARSPM is a viable technique for mapping of 3-D patterns of molecular orientations and LC director dynamics.

In CARS microscopy, a pump wave is used as a probe and the signal is derived by the 3-rd order nonlinear polarization $\sim \chi_{CARS}^{(3)} \vec{E}_p^2(\nu_p) \vec{E}_s(\nu_s)$, where $\vec{E}_p(\nu_p)$ and $\vec{E}_s(\nu_s)$ are the pump and Stokes light waves with frequencies $\nu_p$ and $\nu_s$, respectively; $\chi_{CARS}^{(3)}$ is the 3-rd order nonlinear susceptibility tensor. When vibration resonance condition is satisfied ($\nu_p - \nu_s$ matches a specific Raman band, which is representative of the molecular orientation), the CARS signal at frequency $\nu_{as} = 2\nu_p - \nu_s$ depends on molecular orientations and the image contrast is related to $\hat{n}(\vec{r})$. The polarized probing beams used for CARS allow us to visualize the parts of a sample where $\hat{n}(\vec{r})$ is parallel to $\vec{E}_p(\nu_p) \| \vec{E}_s(\nu_s)$ and/or the linear polarizer in the detection channel; thus similarly to the case of FCPM [2], the images obtained for different CARS polarization states allow one to reconstruct the 3-D pattern of $\hat{n}$.

---
[a] Electronic mail: pnprasad@acsu.buffalo.edu
[b] Electronic mail: Ivan.Smalyukh@Colorado.edu

The experimental set-up is schematically shown in Fig.1. A picosecond Nd:YVO$_4$ *Laser 1* (1064nm, HighQ Laser) with the pulse width ~10 ps and a repetition rate 76MHz is used both as a source of Stokes wave and to synchronously pump *Laser 2*, a tunable (850–890nm) optical parametric oscillator (OPO, HighQ Laser) with the output of ~10ps pulses. The synchronously pumped OPO coherent device provides temporal synchronization with *Laser 1* and serves as a source of the pump wave. Picosecond outputs of *Laser 1* and *Laser 2* were coincided in time and in space and then directed to an inverted microscope. A computer-controlled XY galvano scanner (GSI Lumonics) provided fast scan of the sample in the lateral focal plane of a water-immersion objective *O1* (60X, NA=1.2). The objective *O1* was mounted on a computer-controlled piezo-stage (Piezosystem Jena) for scanning along the microscope's optical axis. Distribution of *Laser1* power between the pump power of *Laser2* and the power of Stokes wave was controlled by the half-wave (λ/2) waveplate *WP1*. Polarizations of *Laser1* and *Laser2* were computer-controlled by rotating Glan-Thomson polarizers *P1* and *P2*. The anti-Stokes CARS signal at $\nu_{as}$ = 722 nm, generated in the forward direction (F-CARS), was collimated by an objective *O2* (NA=0.75) and reflected to a photomultiplier tube (Hamamatsu). The *M10* dichroic mirror and a series of narrow-bandpass barrier filters were used for spectral selection of F-CARS. Polarizer *P3* controlled polarization of the detected F-CARS signal. Backward CARS (E-CARS) and TPF signals were detected in the reflection geometry. Narrow-bandpass barrier filters *F2* and *F3* extracted the respective signals. CARS experiments were performed with parallel orientations of linear polarizations of the input Stokes and pump beams (inset in Fig.1).

LC samples were prepared from glass plates of thickness 0.17 mm, separated by thin (20-40) $\mu m$ mylar spacers and sealed together using an epoxy glue. To align $\hat{n}(\vec{r})$ perpendicular to the glass substrates, we treated the inner surfaces with dilute (<0.05 Wt. %) solutions of surfactant lecithin in hexane. The cells were filled with the room-temperature SmA material, 8CB (octylcyanobiphenyl, Aldrich), by heating the LC to it's isotropic phase (70°C or higher). Some of the samples were doped by a fluorescent dye n, n'-bis(2,5-di-tert-butylphenyl)-3,4,9,10-perylenedicarboximide (BTBP, for the TPF reference experiments) at 0.01Wt.% [2]. Upon cooling from the isotropic phase down to the room temperature, the LC is in the SmA phase with smectic layers parallel to confining plates and $\hat{n}(\vec{r})$ perpendicular to them. Defects in $\hat{n}(\vec{r})$ nucleate during the temperature quench and will be studied using CARS polarized microscopy.

To select a specific vibrational resonance for CARS imaging of 8CB material, we have used a single-spot Raman microspectrometer. We selected the CN vibration in the 8CB molecule (oscillation at 2236 cm$^{-1}$), Fig.2, because its spectral location is well separated from that of other chemical bonds. Using the setup shown in Fig. 1, we have measured high-resolution CARS spectra by tuning $\nu_p$ in the region of $\nu_p - \nu_s$ corresponding to the CN vibration (Fig. 2). The F-CARS signal data are measured for $\vec{E}_p \| \vec{E}_s$ of the input beams parallel to the LC director (shown by solid circles) and for the orthogonal case (open circles). The polarized CARS signal has a much stronger sensitivity to molecular orientations and $\hat{n}(\vec{r})$ than the spontaneous Raman signal for the same vibration band (grey solid and dashed lines in Fig. 2). Moreover, the non-resonant background for selected CN vibration of 8CB is very low.

To demonstrate the 3-D director imaging capabilities of CARSPM, we have selected the so-called "focal conic domains" (FCDs) [1,2], which represent a broad class of equidistance-preserving 3-D configurations of layers in smectic LCs. The polarized CARS imaging visualizes the layers' departures from in-plane orientations in the FCDs, Fig.2. The insets (**i**) and (**ii**) in Fig. 2 show the F-CARS in-plane cross-sections of the FCD at resonance frequency $\nu_p - \nu_s = 2236 cm^{-1}$ for different collinear orientations of $\vec{E}_p \| \vec{E}_s$ and the polarizer in the detection channel. The inset (**iii**) in Fig.2b shows a dramatic intensity degradation of the same-area F-CARS image at $\nu_p - \nu_s = 2252 cm^{-1}$, away from the CN resonance. Image quality depends on the powers of pump and Stokes beams and on the signal integration time. Images could be acquired by raster sample scanning as fast as 500,000 *pixels* per second (integration time $2 \mu s / pixel$), Figs. 2-4, which is ~100,000 times faster than in confocal Raman microscopy [3,4]. The laser powers could be as low as 17mW for the pump wave and 5mW for the Stokes wave. However, even when the used laser powers were an order of magnitude higher, the laser scanning during CARS image acquisition did not alter the director structures in the studied SmA material.

Figure 3 shows in-plane XY sections of a single FCD obtained simultaneously in the modes of (a) F-CARS, (b) E-CARS, and (c) epi-TPF in the LC doped with the BTBP dye. The strongest signal corresponds to the parts of the sample with $\hat{n}(\vec{r})$ parallel to the linear polarization directions of probing light, as shown in Fig.3(d); TPF and CARSPM images in different modes are consistent with our FCPM studies of FCDs [2]. This allows one to map out the pattern of molecular orientation in 3-D, as demonstrated in Fig.4. F-CARS images of a single (Fig.4b) and multiple (Fig.4d) FCDs are constructed from 21 in-plane cross-sections obtained at different depths of sample. In FCDs, the equidistant layers fold around confocal defect lines, the ellipse and hyperbola, Fig.4a. Multiple FCDs of different eccentricity are embedded into the SmA slab with planar stacks of layers, Fig.4d. Experimental images (Fig.4b,d) are consistent with the computer-simulated layered structure (Fig. 4a) and the 3-D director field (Fig. 4c). We have also applied polarized CARS for imaging of structures in a nematic LC pentylcyanobiphenyl (5CB, EM Chemicals) and its cholesteric mixtures (obtained by adding ~0.5Wt.% of chiral agent CB15, EM Chemicals) using oscillation at 2236cm$^{-1}$ due to the CN bond of the molecules and in a ferroelectric SmC* material using vibration at 1608cm$^{-1}$ due to the C=C bond. Thus CARS polarized microscopy has many applications in LC research and can be used to directly probe structures in devises with different types of LCs. The optical anisotropy effects on CARS image contrast and director reconstruction in high- birefringence LCs need to be accounted for and will be discussed in details elsewhere.

To conclude, CARS polarized microscopy is capable of non-invasive 3-D imaging of LC structures. The technique is labeling-free, which eliminates the need of dyes, and enables direct probing of LC director fields in devices and displays. Selective sensitivity to oscillations makes CARSPM especially attractive for imaging of different LC directors in biaxial nematics and smectics by using different chemical bonds of the biaxial LC molecules. Fast image acquisition and short CARS signal integration times might allow one to study temporal dynamics of the director structures.

Research was supported by the Directorate of Chemistry and Life Sciences of Air Force Office of Scientific Research and by International Institute for Complex Adaptive Matter.

1. P.G. de Gennes and J. Prost. *The Physics of Liquid Crystals* (Clarendon Press, Oxford 1993).
2. I.I. Smalyukh, S.V. Shiyanovskii, and O.D. Lavrentovich, Chem. Phys. Lett. **336**, 88 (2001).
3. J.-F. Blach, M. Warenghem, and D. Bormann, Vibr. Spectroscopy **41,** 48 (2006).
4. M. Ofuji, Y. Takano, Y. Houkawa, Y. Takanishi, K. Ishikawa, H. Takezoe, T. Mori, M. Goh, S. Guo, and K. Akagi, Japanese J. Appl. Phys. **45**, 1710 (2006).
5. K. Yoshiki, M. Hashimoto, and T. Araki, Japanese J. Appl. Phys. **44**, L1066 (2005).

6. R.S. Pillai, M. Oh-e, H. Yokoyama, C.J. Brakenhoff, and M. Muller, Opt. Express **14**, 12976 (2006).
7. A. Xie and D.A. Higgins, Appl. Phys. Lett. **84**, 4014 (2004).
8. J.-X. Cheng and X.S. Xie, J. Phys. Chem. B **108**, 827 (2004).
9. J.-X. Cheng, S. Pautot, D.A. Weitz, and X.S. Xie, Proc. Nat. Acad. Sci. USA **100**, 9826 (2003).

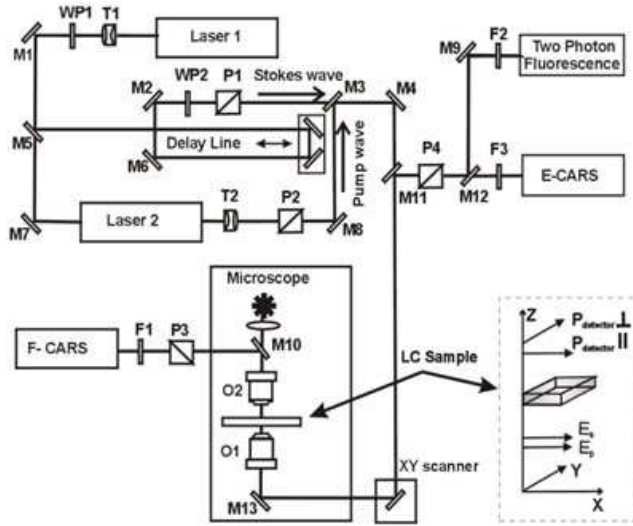

**Fig. 1.** Optical setup of the laser scanning CARS imaging system. *Laser 1* and *Laser 2* are picosecond lasers with the outputs at $\nu_s$ and $\nu_p$, respectively; *T1* and *T2* – lens telescopes; *WP1* and *WP2* – $\lambda/2$ waveplates; *M1 – M13* – dichroic dielectric mirrors; *O1, O2* –objectives; the optical Delay Line, filter wheels *F1 – F3,* Glan-Thompson polarizers *P1 – P4,* and the galvano *XY* scanner are all controlled by a computer; photomultiplier tubes in the channels of TPF, E-CARS and F-CARS are marked respectively.

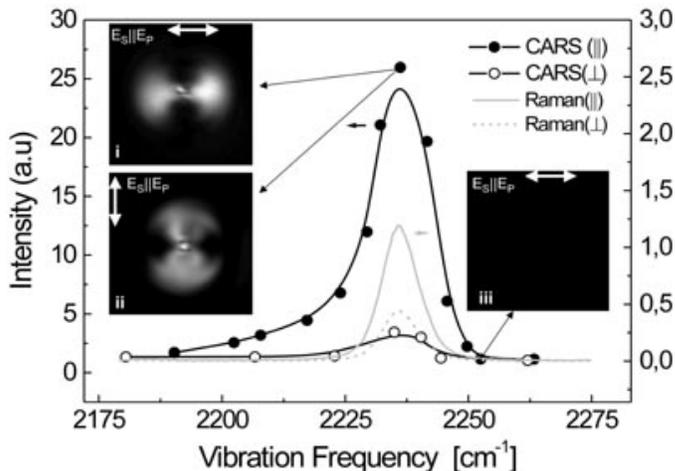

**Fig. 2.** CARS and Raman spectral lines of the CN bond of 8CB for laser polarizations parallel and perpendicular to $\hat{n}$. The insets (i), (ii) and (iii) show F-CARS in-plane images of the FCD at (i, ii) resonance frequency 2236 cm$^{-1}$ for two orthogonal polarizations of detected CARS signal, and (iii) at frequency 2252 cm$^{-1}$ (away from the resonance). Image size in the insets is 300 x 300 pixels corresponding to the area 21 x 21 μm.

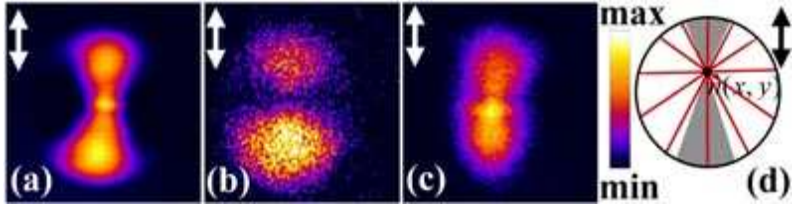

**Fig. 3.** Depth-resolved single-scan cross-sections of the FCD obtained in nonlinear microscopy modes of (a) F-CARS; (b) E-CARS; (c) epi-TPF. Image size is 12.6×12.6µm (300×300 pixels). (d) FCD's director structure in the plane of ellipse: $\hat{n}(x, y)$ is in-plane radial inside of the ellipse and vertical $\hat{n} \| \hat{z}$ outside; dark bands mark regions of the strongest CARSPM signal, where $\hat{n}$ is parallel to linear polarizations of probing beams.

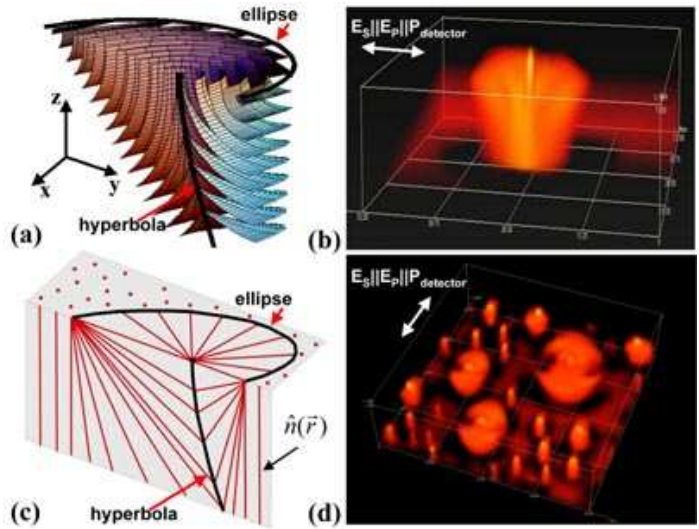

**Fig. 4.** FCDs in a stack of parallel smectic layers: (a) cross-section of the layered structure and (c) corresponding 3-D director field with the marked hyperbola and ellipse defects; (b,d) 3-D CARSPM images of $\hat{n}(\vec{r})$ in samples with (b) single FCD and (d) multiple FCDs embedded into flat parallel layers; the images have been reconstructed using a series of XY cross-sections obtained in the F-CARS mode at different depths of sample and with 1µm step in Z-direction. The sample area is 53x53µm in (b) and 64x64µm in (d).